\documentclass{PoS}

\title{Powerful extragalactic jets} 

\ShortTitle{Powerful extragalactic jets}

\author{\speaker{Gabriele Ghisellini} \\
        INAF -- Osservatorio Astronomico di Brera\\
        E-mail: \email{gabriele.ghisellini@brera.inaf.it}}

\abstract{
The {\it Fermi}, {\it Swift} and {\it INTEGRAL} satellites, together with ground based 
(especially Cherenkov) telescopes made possible a great progress in our understanding 
of relativistic jets. We can now start to attack the difficult questions of jet 
formation, collimation and content. We can also used them as probes to quantify the 
amount of IR and optical background radiation, and the amount of the cosmic magnetic field. 
Since they are the most powerful steady sources of the Universe, we can study them also 
at large redshifts, and this is a very fruitful field of research. 
To this aim, I will emphasize the importance of high energy X--rays, 
where very powerful blazars are predicted to emit most of their electromagnetic power.
For them, the emission of the underlying accretion disk becomes unhidden
by the non--thermal jet radiation, allowing to estimate the black hole mass
and the accretion rate.
In turn, this highlights the connection between the disk and the jet.
Since the highest power blazars could have their emission peak in the 
$\sim$MeV band, hard X--ray instruments could be more appropriate than 
the {\it Fermi}/LAT to detect them.
}

\FullConference{The Extreme and Variable High Energy Sky - extremesky2011,\\
		September 19-23, 2011\\
		Chia Laguna (Cagliari), Italy}

\begin{document}

\section{Introduction}

BL Lacs and Flat Spectrum Radio Quasars (FSRQs), collectively called blazars, 
have relativistic jets that point at us. 
Due to relativistic beaming, their flux is enhanced and they can therefore
be visible up to large redshifts.
This makes these sources good probes for studying the physics
of jets and to explore some interesting properties of
the far Universe.
Their spectral energy distribution (SED) is always characterized 
by two broad humps (in $\nu F_\nu$), the first peaking at mm to UV frequencies,
the second peaking in the MeV--GeV (and sometimes TeV) bands \cite{fossati98}.
While the origin of the first peak is certainly due to synchrotron,
there is some debate about the origin of the high energy peak:
the prevalent hypothesis is that it is due to the same electrons responsible for
the synchrotron peak, scattering their own synchrotron photons in
low power BL Lac sources (SSC, \cite{maraschi92}), and scattering radiation produced 
externally to the jet (EC) in high power FSRQs 
\cite{dermer93}, \cite{sikora94}, \cite{ghisellini98}.
Through simultaneous data covering the IR to $\gamma$--ray band, we can now derive
several interesting parameters of the jet emitting region, and, in very powerful
sources, we directly see the contribution of the accretion disk in the
optical band. 
Thus, the black hole mass and accretion rate can be estimated, allowing to
compare accretion and jet powers, both in absolute terms and when these
quantities are measured in Eddington units.

The {\it Fermi} satellite allowed a huge jump in strengthening the knowledge of
the SED of blazars since the {\it Compton Gamma Ray Observatory} era,
detecting several hundreds of blazars of all kinds.
But the added value of X--ray observations is also huge:
hard X--rays above 10 keV in blazars are particularly important at the two extremes of
the so--called ``blazar sequence" \cite{fossati98}: i) in low power BL Lacs
they can be due to the tail of the synchrotron spectrum, making them
good candidates as strong TeV emitters: ii) on the high power end of the sequence, 
namely in very powerful FSRQs, the hard X--ray flux is close to the emission peak, that 
in these sources is in the MeV energy range, and is dominating the bolometric output.
Therefore in these sources hard X--rays carry a very significant fraction 
of the jet luminosity, making them visible and detectable at
very high redshift.

This poses the question: to find out the most powerful blazars at high redshift,
what is the best energy band and instrument?
Hard X-rays (thus INTEGRAL and {\it Swift}/BAT) or $\gamma$--rays (i.e. {\it Fermi}/LAT)?
The answer of course depends on the average source flux in the two bands coupled
with the corresponding sensitivity.
I will here argue that hard X--rays are more promising.

\begin{figure}
\vskip -0.4 cm
\center
\includegraphics[width=.65\textwidth]{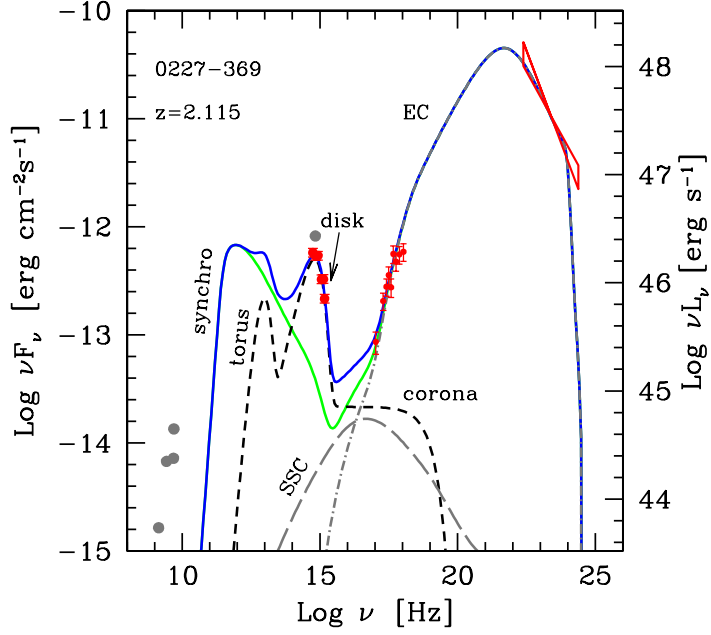}
\vskip -0.7 cm
\caption{
The SED blazar 0227--369 from the radio to the $\gamma$--rays. The lines refer to 
the leptonic, one--zone model used in \cite{ghisellini09}.
The different components are labelled. 
Note the synchrotron jet continuum that peaks in the mm band and is steep 
[$\alpha>1$; $F(\nu) \propto \nu^{-\alpha}$] after the peak.
This leaves the accretion disk contribution unhidden. By modelling the disk component
we can derive the black hole mass and accretion rate.
}
\label{0227}
\end{figure}

\section{Naked disks in high power FSRQs}

The main distinguishing feature between BL Lacs and FSRQs is the presence or
absence of the broad emission lines. 
In FSRQs they are well visible, and flag the presence of an ionizing
continuum, produced by an accretion disk.
The synchrotron hump in these powerful FSRQs peaks in the far IR and mm band,
and is steep after the peak [namely, $\alpha>1$, with $F(\nu) \propto \nu^{-\alpha}$].
This is confirmed by the slope of the $\gamma$--ray flux, as
detected by {\it Fermi}.
Furthermore, in these sources the synchrotron component is relatively weak
with respect to the high energy one.
Fig. \ref{0227} shows a typical example of the SED of these powerful blazars.
It can be seen that the location of the synchrotron peak leaves the contribution
of the accretion disk unhidden, and therefore well visible.

We can fit it by applying for instance a simple Shakura--Sunyaev \cite{shakura73}
model, and find both the black hole mass and the accretion rate.
A posteriori, we can also check if the disk luminosity is above $10^{-2} L_{\rm Edd}$,
justifying the use of a Shakura--Sunyaev disk. 
For BL Lacs, instead, there is no direct sign in the SED of the accretion disk.
However, in a number of them, broad lines, albeit weak, have been detected
and for several others we \cite{sbarrato11} could provide an upper limit.
Then, by using the template of  \cite{francis91}, we could estimate the luminosity
of the entire Broad Line Region (BLR), and compare it to the $\gamma$--ray luminosity.

\section{Broad lines and $\gamma$--rays}

\begin{figure}
\vskip -0.4 cm
\center
\includegraphics[width=.65\textwidth]{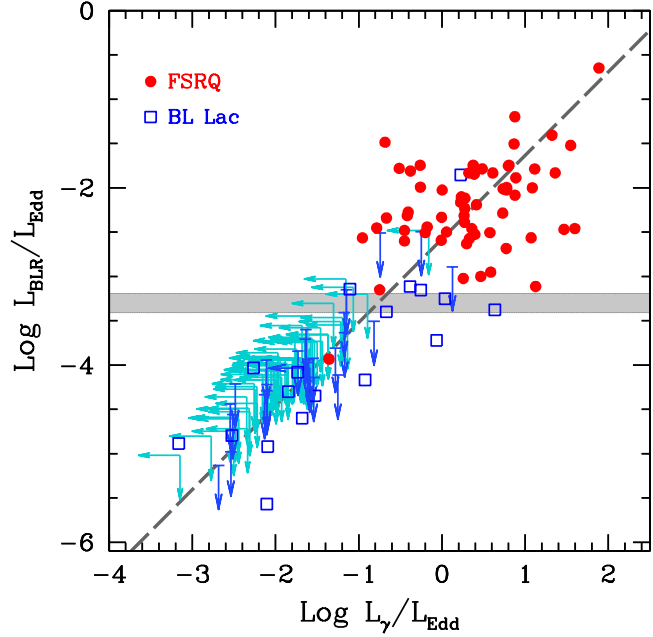}
\vskip -0.7 cm
\caption{The luminosity of the broad emission lines as a function of
the $\gamma$--ray luminosity as detected by {\it Fermi}.
Both are in Eddington units. 
Red and blue symbols correspond to FSRQs and BL Lacs, respectively. 
Arrows corresponds to upper limits.
The correlation (dashed line) is almost linear.
The grey horizontal stripe indicates $L_{\rm BLR}/L_{\rm Edd}=5\times 10^{-4}$,
which best divides BL Lacs and FSRQs (as classified as such on the
base of the equivalent widths of their lines).
Adapted from \cite{sbarrato11}. 
}
\label{blrgamma}
\end{figure}

Fig. \ref{blrgamma} shows the correlation between the BLR and the $\gamma$--ray luminosities, 
both in Eddington units (but a strong correlation is present also when considering  the absolute quantities).
This figure, adapted from \cite{sbarrato11}, shows also how BL Lacs and FSRQs divide,
at a luminosity $L_{\rm BLR}/L_{\rm Edd}\sim 5\times 10^{-4}$.
This is the value that better separates the two classes of blazars.
We have proposed to adopt this division when classifying blazars, since it is more
physical that the classical classification on the base of the equivalent width of
the broad emission lines.
Since the BLR luminosity is associated with the disk luminosity, and
the $\gamma$--ray one is associated with the jet power (if the 
viewing angle and bulk Lorentz factor are similar for all blazars),
then Fig. \ref{blrgamma} shows a clear link between the jet power and 
the accretion luminosity. 
Since a specific source can vary its $\gamma$--ray luminosity by even
two orders of magnitude, (see for instance \cite{ghirlanda11}), we
should not be surprised by the large scatter around this correlation.
If the BLR, on average, intercepts and re--emits $\sim 1/20$
of the disk luminosity $L_{\rm d}$, then the value $L_{\rm BLR}/L_{\rm Edd}=5\times 10^{-4}$ 
corresponds to $L_{\rm d}/L_{\rm Edd}\sim 10^{-2}$.
This can correspond to the transition between a standard and a radiatively
inefficient accretion regime. 
But there is another possibility, suggested by the linearity of the observed 
$L_{\rm BLR}$--$L_\gamma$ correlation (although the paucity of points
cannot allow any robust claim): even if the radiatively inefficient/efficient 
transition happened at much lower values of $L_{\rm d}/L_{\rm Edd}$
(as suggested in \cite{sharma07}),
the relation between the size of the BLR and $L_{\rm d}$ implies
very small BLR sizes when $L_{\rm d}$ is small. 
If the dissipation region is instead always a multiple of the Schwarzschild radius
(about a thousand), objects with weak lines would have jets dissipating
and producing most of their radiation {\it outside} the BLR.
In this case the EC process would be not important even if 
the broad lines are indeed produced \cite{sbarrato11}.

\begin{figure}
\vskip -0.4 cm
\center
\includegraphics[width=.65\textwidth]{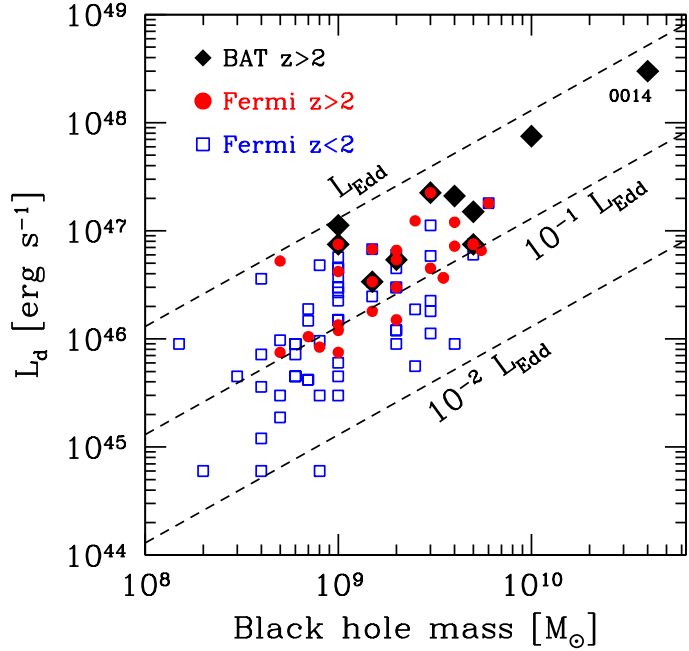}
\vskip -0.5 cm
\caption{
Accretion disk luminosity $L_{\rm d}$
as a function of black hole mass for blazars with $z>2$ in the
BAT sample (diamonds; A09) and in the 1LAC {\it Fermi}/LAT sample
(circles, see \cite{abdo1LAC}).
Empty squares are FSRQs in 1LAC at $z<2$ (\cite{ghisellini09}, 
\cite{ghisellini10} and \cite{ghisellini11}).
All FSRQs have $L_{\rm d}/L_{\rm Edd}>10^{-2}$, and
all high redshift BAT blazars have black holes
with $M>10^9 M_\odot$ and $L_{\rm d}/L_{\rm Edd}>0.1$.
}
\label{blazars}
\end{figure}

\section{Black hole masses and accretion rates}

Fig. \ref{blazars} shows the accretion disk luminosity $L_{\rm d}$
as a function of the black hole mass for all FSRQs analyzed in 
\cite{ghisellini09}, \cite{ghisellini10} and \cite{ghisellini11}.
These values have been derived by fitting the optical--UV
data with a standard disk.
The presence of the synchrotron component in some sources, while is
accounted for by the fit, inevitably introduces some uncertainties
when it is strongly contributing to the optical--UV continuum,
but the presence of the broad emission lines in any case
allows to estimate the luminosity of the disk in a relatively
accurate way (the uncertainties here being the reconstruction
of the entire BLR luminosities on the base of one or two lines,
and the BLR covering factor).
It can be seen that all $\gamma$--ray loud FSRQs we have studied have
$L_{\rm d}/L_{\rm Edd}>10^{-2}$, and are therefore in the radiatively
efficient regime of accretion. 
This, a posteriori, justifies using the standard accretion disk 
as a fitting model.
Three group of sources are shown: the {\it Fermi} detected FSRQs at $z<2$,
those at $z>2$, and the FSRQs detected by BAT at $z>2$.
The latter are the most powerful, in Eddington units.
All FSRQs at $z>2$ detected by BAT have black hole masses $M>10^9 M_\odot$
and disks emitting at more than 10\% of the Eddington limit.
They appear more extreme than the high redshift FSRQs detected by {\it Fermi}.

\subsection{The case of S5 0014+813}

The source with the largest disk luminosity and black hole mass
is S5 0014+813, at $z=3.366$.
In \cite{massona09} we have derived a black hole mass as ``outrageous" as $M=4\times 10^{10}M_\odot$
accreting at 40\% Eddington, thus producing a disk luminosity $L_{\rm d}\sim 2\times 10^{48}$
erg s$^{-1}$, which is what observed in the NIR-optical--UV.
Discussing this case, we have proposed a solution that would allow to have a smaller
black hole mass, i.e. that the disk radiation is collimated (i.e. {\it not} beamed)
by a funnel.
If the solid angle of the funnel is $\Delta\Omega_{\rm funnel}$, one can reduce the power budget
by $\Delta\Omega_{\rm funnel}/4\pi$.
Since this source is a blazar, the viewing angle with respect to the jet axis
is small, ad this ensures that we are looking down to the funnel,
since the axis of the funnel and the axis of the jet likely coincide.
This would easily allow for a factor $\sim10$ of apparent amplification of the 
accretion disk flux, and then we could reduce the required black hole mass
by an order of magnitude.

On the other hand, the broad emission lines are also powerful, with a Lyman--$\alpha$
luminosity of $\sim 10^{46}$ erg s$^{-1}$ \cite{sargent89}, and this emission is surely isotropic.
The ionizing continuum cannot have a ``true" luminosity smaller than 
$\sim 10 \times L_{\rm BLR} \sim 5\times 10^{47}$ erg s$^{-1}$.
If the ionizing luminosity coincides with the entire accretion luminosity,
then we would require $M>4\times 10^9 M_\odot$, to be sub--Eddington.
But with such a large mass the ionizing luminosity is only a fraction of the 
entire disk emission (i.e. about 10\%), so we require more power, and thus
an heavier black hole, making the value $M=4\times 10^{10}M_\odot$ inevitable
(within a factor 2).

\begin{figure} 
\begin{tabular}{cc}
\includegraphics[width=.5\textwidth]{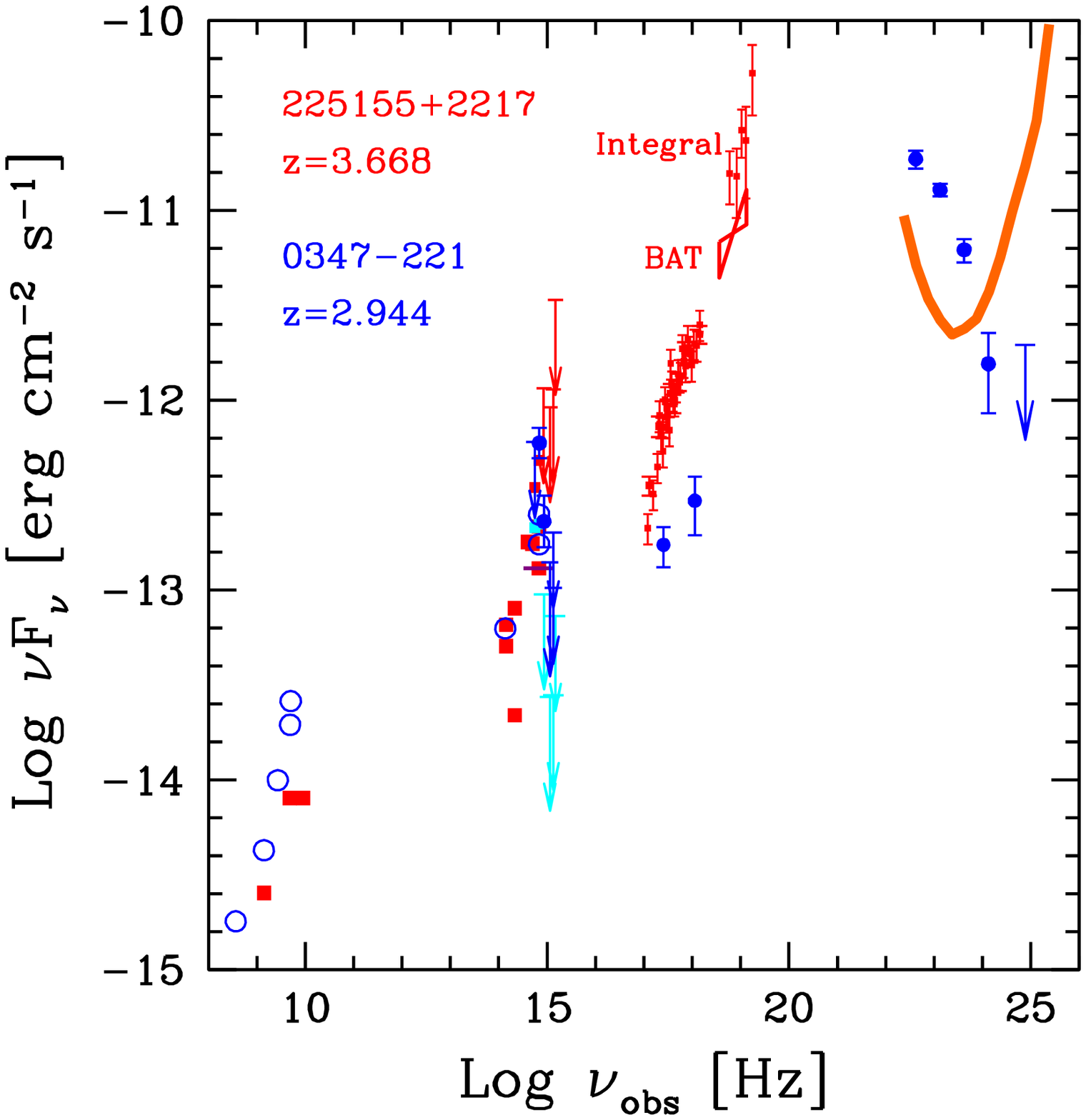} &
\includegraphics[width=.5\textwidth]{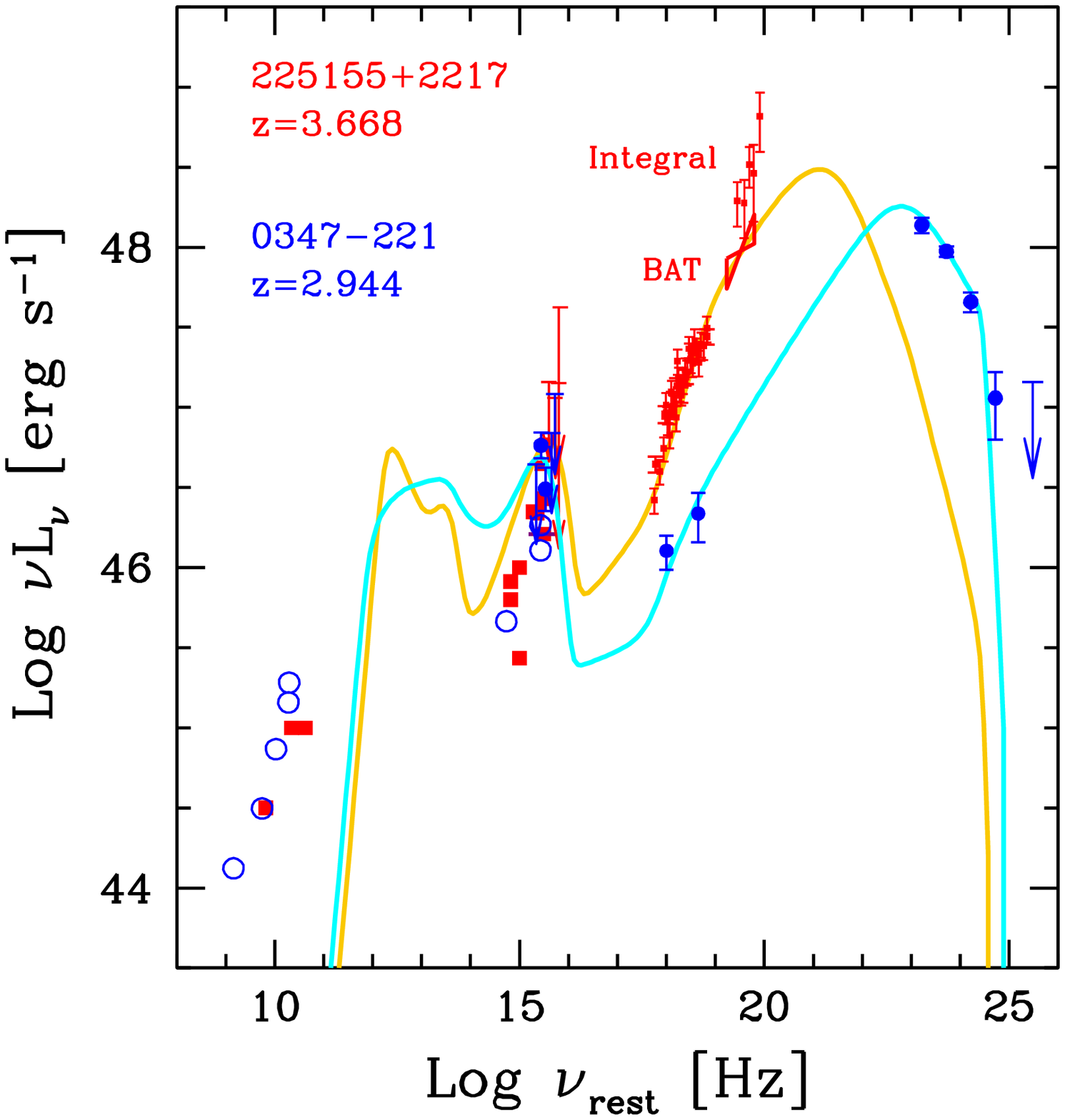} 
\end{tabular}
\vskip -0.5 cm
\caption{
Comparison between the SED of the two more distant blazars detected by BAT (225155+2217; $z$=3.668)
and by LAT (0347--221; $z$=2.944). 
On the left we plot the $\nu F_\nu$ flux vs the observed frequency, on the right we plot $\nu L_\nu$ vs the
rest frame frequency.
We also plot a single--zone leptonic model that fits the 
sources. (From \cite{ghisellini09} and \cite{ghisellini10}). The orange line in the left panel
is the limiting 5$\sigma$ sensitivity of LAT after 1 year of survey.  For
a more detailed analysis of the INTEGRAL data of 225155+2217 see \cite{lanzuisi11}.  } 
\label{2sed} 
\end{figure}

\section{Fermi/LAT vs Swift/BAT}

The 3--year survey of {\it Swift}/BAT detected 38
blazars in the [15--55 keV] band.
Of these, 10 are at $z>2$, and 5 of them are at $z>3$.
All of these high redshift blazars have luminosities $L_X >2 \times 10^{47}$ erg s$^{-1}$.
A recent update using the 58 months survey \cite{baum10} brings the number
of $z>2$ blazars to 16, 6 of which are at $z>3$.
We can compare these numbers with the total number of blazars detected by {\it Fermi}
at $z>2$:  these are 28 (with only 2 at $z>3$) in the ``clean" 1LAC catalog \cite{abdo1LAC},
and 31 (with 2 at $z>3$) in the ``clean" 2LAC sample \cite{ackermann11}
(note that some blazars in the 1LAC sample are not present in 2LAC, because their
average 2--year flux went below the assumed flux threshold).
We then conclude that both in absolute and especially in relative terms the hard X--ray 
observations are more efficient than $\gamma$--ray ones to select blazars at high redshifts.
Fig. \ref{2sed} illustrates this case, by comparing two high--$z$ blazars: one (225155+2217; $z$=3.668)
has been detected by BAT (and not by LAT), the other (0347--221; $z$=2.944) has
been detected by LAT (and not by BAT).
225155+2217 is more powerful, its high energy peak is at $\sim$1 MeV, and it is much more powerful in
hard X--rays than 0347--221, whose high energy peak should be located at larger energies.

\begin{figure} 
\center
\includegraphics[width=.65\textwidth]{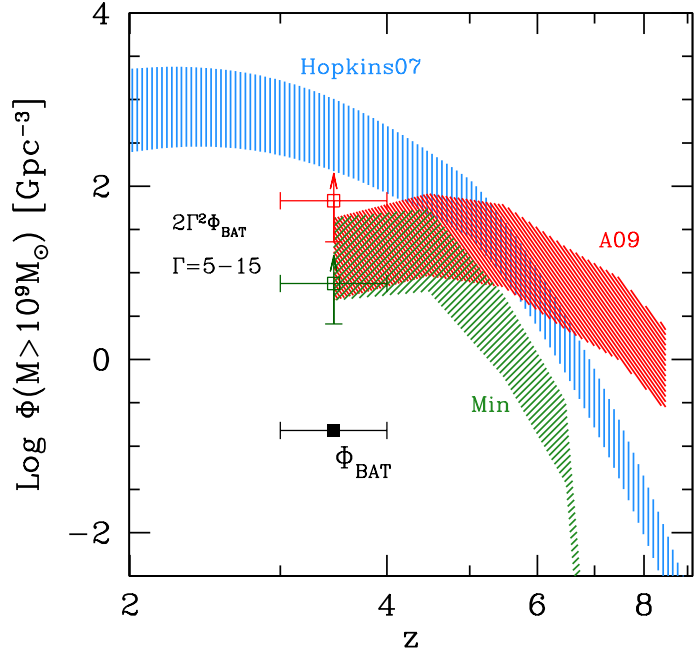} 
\vskip -0.7 cm
\caption{
The mass function of black holes with masses $M>10^9 M_\odot$.
The black square in the redshift bin $3<z<4$ is the value
considering only FSRQs in the 3 years BAT survey \cite{ajello09}.
Multiplying it by $2\Gamma^2$ we have the green ($\Gamma=5$) or red ($\Gamma=15$)
points.
The stripes are the extrapolation to larger redshifts of the BAT blazar luminosity function,
according to the evolution proposed by \cite{ajello09} and to the minimal evolution
discussed in \cite{ghisellini10}.
The blue stripe corresponds to the mass function of radio--quiet objects (with 
optical luminosities larger than $10^{47}$ erg s$^{-1}$), \cite{hopkins07}.
Adapted from \cite{volonteri11}.  } 
\label{fi} 
\end{figure}

\section{Heavy early black holes}

All FSRQs at $z>2$ detected by BAT during the first 3 years \cite{ajello09}
have an estimated black hole mass exceeding $10^9 M_\odot$ \cite{ghisellini10}.
These are also those FSRQs exceeding a luminosity of $L_X=2\times 10^{47}$ erg s$^{-1}$
in the [15--55 keV] band. 
Therefore the luminosity function above this value of luminosity directly 
gives a lower limit on the mass density of black holes, {\it in blazars}, with $M>10^9 M_\odot$.
Fig. \ref{fi} shows this estimate as a black square (labelled $\phi_{\rm BAT}$),
in the $3<z<4$ redshift bin.
But the real density of these heavy black holes is a factor $2\Gamma^2$ higher, 
where $\Gamma$ is the bulk Lorentz factor of the X--ray emitting jet.
Therefore Fig. \ref{fi} shows the density of heavy black holes multiplying what
directly derived for blazars by a factor 50 (i.e. $\Gamma=5$) or 450 ($\Gamma=15$).
Then, assuming the luminosity function of \cite{ajello09} and its extrapolation above
$z=4$ (where we have no data), we have the red stripe (labelled A09).
The green stripe, instead, (labelled as Min), corresponds to a different evolution 
of the luminosity function of \cite{ajello09}, but only above $z=4$.
It is a ``minimal" luminosity function because it is consistent with the
few powerful blazars already detected at $z>4$ (and with $M>10^9M_\odot$), 
discovered serendipitously \cite{ghisellini10}, \cite{volonteri11}.
The two mass functions are then equal for $z<4$, but become quite different above.
We can then compare them with the mass function of heavy black holes in
radio--quiet quasars.
To this aim Fig. \ref{fi} shows the one derived taking the luminosity function
of \cite{hopkins07}, and integrating the density of objects above $L=10^{47}$ erg s$^{-1}$
in the optical (i.e. masses above $10^9M_\odot$, if they are Eddington limited).
This is shown by the blue stripe (labelled Hopkins07).
In \cite{volonteri11} we have then stressed that the luminosity function of \cite{ajello09}
yields a density of heavy and early black holes of radio--loud objects 
that is larger than what derived for radio--quiet ones.
Even if strange, this is not impossible: it could be that, to form a very massive 
black hole in a short time (i.e. high $z$) the system {\it requires} a jet.
On the other hand, there is a more conservative solution, depicted by the ``minimal"
mass function, where the factor $\sim 1/10$ of the ratio between radio--loud and radio--quiet
is maintained also at large $z$.
Finding the true mass function at large $z$ of blazars is the next challenge.

\end{document}